# Lack of simple correlation between switching current density and spin-orbit torque efficiency of perpendicularly magnetized spin-current generator/ferromagnet heterostructures

Lijun Zhu[1*], D. C. Ralph[1,2], R. A. Buhrman[1]
1. Cornell University, Ithaca, New York 14850, USA
2. Kavli Institute at Cornell, Ithaca, New York 14850, USA
*lz442@cornell.edu

Spin-orbit torque can drive electrical switching of magnetic layers. Here, we report that at least for micrometer-sized samples there is no simple correlation between the efficiency of dampinglike spin-orbit torque ($\xi_{DL}^j$) and the critical switching current density of perpendicularly magnetized spin-current generator/ferromagnet heterostructures. We find that the values of $\xi_{DL}^j$ based on switching current densities can either under- or over-estimated $\xi_{DL}^j$ by up to tens of times in a domain-wall depinning analysis, while in the macrospin analysis based on the switching current density $\xi_{DL}^j$ can be overestimated by up to thousands of times. When comparing the relative strengths of $\xi_{DL}^j$ of spin-current generators, the critical switching current densities by themselves are a poor predictor.

Strong dampinglike spin-orbit torque (SOT) can efficiently switch the magnetic free layers of magnetic tunnel junctions (MTJs) at the nanosecond timescale [1-5]. Enormous efforts have been made on developing strong spin current generators (SCGs)[6-10] and high spin-transparency interfaces [11-15] that can provide high dampinglike SOT efficiency ($\xi_{DL}^j$). This is mainly motivated by the fact that $\xi_{DL}^j$ of a SCG/ferromagnet (FM) heterostructure directly connects to the density of the critical switching current inside the SCG layer ($j_c$) and thus the *total* switching current ($I_c$, the sum of the currents in the SCG and the FM layers) that will define the energy efficiency ($\propto I_c^2$), the scalability [16](the transistor dimension $\propto I_c$), and the endurance (electro-immigration $\propto I_c^2$ [17]) of SOT-MTJs. In simplified models, $\xi_{DL}^j$ of a heterostructure with perpendicular magnetic anisotropy (PMA) inversely correlates to $j_c$ via Eq. [1] in the macrospin limit [18-20] and via Eq. [2] in the domain wall depinning regime [21-23],

$$j_c = e\mu_0 M_s t_{FM} (H_k - \sqrt{2}|H_x|)/\hbar\xi_{DL}^j, \quad (1)$$
$$j_c = (4e/\pi\hbar)\,\mu_0 M_s t_{FM} H_c/\xi_{DL}^j, \quad (2)$$

where $e$ is the elementary charge, $\hbar$ is the reduced Planck constant, $\mu_0$ is the permeability of vacuum, $H_x$ is the applied field along the current direction, and $t_{FM}$, $M_s$, $H_k$, and $H_c$ are the thickness, the saturation magnetization, the effective perpendicular anisotropy field, and the perpendicular coercivity of the FM, respectively.

While the main mechanism of perpendicular magnetization switching in micrometer-sized devices [24-26] is well understood to be domain wall depinning rather than coherent macrospin rotation, it has still remained open questions as to how accurate Eq. [2] can quantitatively describe real SCG/FM systems and what level of discrepancy a simplified macrospin analysis can result in. Moreover, it has also remained unclarified as to whether the $j_c$ values can provide a guidance about the relative $\xi_{DL}^j$ strengths of different SCG/FM samples. Those open questions worth a systematic quantitative study because it would be highly desirable if $\xi_{DL}^j$ can be determined or predicted from switching currents.

In this work, we investigate in detail the quantitative correlation between $\xi_{DL}^j$ and $j_c$ in a variety of perpendicularly magnetized SCG/FM samples. We show that there is no simple correlation between $\xi_{DL}^j$ and $j_c$ at least for micrometer-sized perpendicular SCG/FM samples. We find that quantitative analyses of the switching current densities based on Eq. [1] or Eq. [2] can result in orders of magnitude errors compared to the results from harmonic Hall voltage response (HHVR) analysis of current-induced small-angle magnetization tilting. These results indicate that the current-induced switching of perpendicular magnetization cannot predict quantitative results of the dampinglike SOT.

As listed in Table 1, we sputter-deposited eight perpendicularly magnetized SCG/FM bilayers: Pt 2/Co 1.4 (annealed in vacuum at 300 °C for 1 hour), Pt 4/Co 0.63, Pt-Hf/Co 0.75 (Pt-Hf = [Pt 0.6/Hf 0.2]$_5$/Pt 0.6 multilayers), Pt$_{0.75}$Pd$_{0.25}$ 4/Co 0.64, Au$_{0.25}$Pt$_{0.75}$ 4/Co 0.64, Pt$_{0.7}$(MgO)$_{0.3}$ 4/Co 0.68, Pd 4/Co 0.64, and W 4/Fe$_{0.6}$Co$_{0.6}$B 1.5 (annealed in vacuum at 350 °C for 1 hour). Here, the layer thicknesses are in nm. Each bilayer was deposited on an oxidized Si substrate with a 1 nm Ta seed layer and was protected by a MgO 2/Ta 1.5 bilayer, the latter is fully oxidized after exposure to atmosphere (see the electron energy loss spectrum results in Ref. [27]). The values of $M_s$ for the FM layers are measured by vibrating sample magnetometry. We determined the resistivity ($\rho_{xx}$) of the SCG layers (Table 1) by measuring the conductance enhancement of the corresponding stacks with respect to a reference stack with no SCG layer. The samples were patterned into Hall bars that are 5 μm in width and 60 μm in length ($L$) for HHVR and switching experiments using the geometry in Fig. 1(a). During the current switching experiments, a dc current was also sourced to the Hall bars by a Keithley 2400 source-meter.

We first clarify that the magnetic field switching of our samples do behavior distinctly from a macrospin and agrees better with the expectation of domain wall depinning. If the thin FM followed Stoner-Wohlfarth macrospin behavior, $H_c$ should be close to $H_k$ at the field polar angle of $\theta_H = 0°$ in any finite measurement time, and $H_c$ should vary with $H_k$ ($\cos^{2/3}\theta_H + \sin^{2/3}\theta_H$)$^{-3/2}$ [28]. However, the observed $H_c$ for both SCG/Co and SCG/FeCoB samples is much smaller than $H_k$ and significantly deviates from the scaling $H_k$ ($\cos^{2/3}\theta_H + \sin^{2/3}\theta_H$)$^{-3/2}$ [see Fig. 1(b) and 1(c)]. Instead, $H_c$ much more closely follows a $1/\cos\theta_H$ scaling, which is predicted by the switching via thermally-assisted reversed domain nucleation and domain wall propagation [24,29]. For the Pt/Co sample, there is some deviation from the $1/\cos\theta_H$ behavior as $\theta_H$ approaches 90° most likely due to coherent rotation of the magnetization vector in the pinned domain when the in-plane hard-axis field component is sufficiently strong [24,29]. Note that such deviation is absent in the W 4/FeCoB 1.5 sample with a weak pinning field ($H_c$ = 40 Oe), suggesting minimal magnetization rotation in this sample even when $\theta_H$ is fairly close to 90°. These observations are consistent with previous perpendicular MTJ experiments that the rigid macrospin reversal *never* happens unless the device size is rather small (< 50 nm)[5,30].

It has been established that the HHVR measurements can provide reliable determination of $\xi_{DL}^j$ of both in-plane and perpendicularly magnetized SCC/FM heterostructures and yield results that are quantitatively consistent with those obtained from shifts of out-of-plane switching fields of PMA SCC/FM heterostructures due to current-induced effective field on the Néel domain walls [26] (see below). Therefore, as a standard against which to evaluate the switching current means of estimating $\xi_{DL}^j$, we first determine $\xi_{DL}^j$ for all the samples using the "out-of-plane" and/or angle-dependent "in-plane" HHVR measurements [31-33], with the values listed in Table 1. We apply a sinusoidal voltage ($V_{in}$) to the current leads of the Hall bar (which is along $x$ direction) and measure the in-phase first and the out-of-phase second harmonic Hall voltages, $V_{1\omega}$ and $V_{2\omega}$, using a lock-in amplifier. For "out-of-plane" HHVR measurement on PMA samples [Fig. 2(a)], the dampinglike effective SOT field ($H_{DL}$) is given by [31,32]

$$H_{DL} = -2\frac{\partial V_{2\omega}}{\partial H_x}/\frac{\partial^2 V_{1\omega}}{\partial H_x^2}, \quad (3)$$

where $V_{1\omega}$ and $V_{2\omega}$ are parabolic and linear functions of $H_x$ [Fig. 2(b)], respectively. With the ratio of $H_{DL}$ and the current density ($j = V_{in}/L\rho_{xx}$) in the HM [Fig. 2(c)], $\xi_{DL}^j$ can be determined as

$$\xi_{DL}^j = 2e\mu_0 M_s t_{FM} H_{DL}/\hbar j. \quad (4)$$

Here we do not apply the so-called "planar Hall correction" in analyzing the out-of-plane HHVR results for the reasons discussed in detail in the Supplementary Materials of Refs. [14] and [32]. We have also checked these PMA HHVR results against angle-dependent in-plane HHVR measurements [33] on corresponding samples in which the magnetic layer has in-plane anisotropy due either to being thicker or to being un-annealed, and in all cases we find good consistency. The results for an un-annealed in-plane magnetized Pt 2/Co 1.4 ($M_s$ = 1140 emu/cm$^3$, $\rho_{xx}$ = 57.5 μΩ cm) are shown in Figs. 2(d)-2(f), which agree with $\xi_{DL}^j \approx 0.15$ for the annealed PMA bilayer of Pt 2/Co 1.4 ($M_s$ = 1454 emu/cm$^3$, $\rho_{xx}$ = 67.1 μΩ cm). In the in-plane HHVR measurement, the magnetization remains saturated and rotates following the large in-plane bias field $H_{xy}$. As shown in Fig. 2(d), $V_{2\omega}$ follows

$$V_{2\omega} = (V_{DL} + V_{ANE})\cos\varphi + V_{FL}\cos\varphi\cos2\varphi, \quad (5)$$

where $V_{DL} = -V_{AH}H_{DL}/2(H_{xy}+H_k)$ is the second HHVR to the dampinglike torque, $V_{FL}$ is the second HHVR to the fieldlike torque (including Oersted field torque), $V_{ANE}$ is the anomalous Nernst voltage, $H_{xy}$ is the in-plane bias field, $\varphi$ is the in-plane angle of $H_{xy}$ and thus the magnetization with respect to the current direction. We determine $H_{DL}$ for each $V_{in}$ from the slope of the linear fit of $V_{DL}$ vs - $V_{AH}/2(H_{xy}+H_k)$ [see Figs. 2(e) and 2(f)], and then estimate $\xi_{DL}^j$ using Eq. (4). The corresponding results for angle-dependent in-plane HHVR measurements on in-plane stacks with thicker Co are available in ref. [34] for Pt 4/Co, Pt$_{0.75}$Pd$_{0.25}$ 4/Co, and Pd 4/Co, ref. [35] for Pt-Hf/Co, ref. [36] for Au$_{0.25}$Pt$_{0.75}$ 4/Co, ref. [37] for Pt$_{0.7}$(MgO)$_{0.3}$ 4/Co. For one sample, W 4/Fe$_{0.6}$Co$_{0.2}$B$_{0.2}$ 1.5, $V_{1\omega}$ did not show well-defined parabolic scaling even in the very small $H_x$ regime, despite the fact that the sample has PMA (which is consistent with the absence of any macrospin rotation behavior at around $\theta_H \approx 90°$, see Fig. 1(b)). For this sample, we only obtained the value of $\xi_{DL}^j$ from angle-dependent in-plane HHVR measurement on a control W 4/Fe$_{0.6}$Co$_{0.2}$B$_{0.2}$ 1.8 bilayer that underwent the same post-annealed treatment as the W 4/Fe$_{0.6}$Co$_{0.2}$B$_{0.2}$ 1.5 sample.

We note that the values of $\xi_{DL}^j$ we obtain from the HHVR measurements for all these perpendicular bilayers are all below 0.4. $\xi_{DL}^j$ is 0.14 for the Pt 2/Co 1.4 and 0.21 for Pt 4/Co 0.63, which agrees well with previous HHVR reports from different groups [38,39]. To expand our analysis, Table 1 also includes two results from the literature: for Pt 6/Fe$_3$GeTe$_2$ 4 from ref. [40] and Ta 5/Tb$_{20}$Fe$_{64}$Co$_{16}$ 1.8 from ref. [41]. Here, we do mean to imply that there is not anything incorrect about the analyses in these two papers; their results are convenient for our purposes because their HHVR measurements produced reasonable values for $\xi_{DL}^j$ and they also provided measurements of switching currents.

We now consider to what extent one might accurately estimate $\xi_{DL}^j$ by simply measuring the critical current density for switching of magnetic bilayers with PMA and applying the domain wall effective field model (Eq. [2]) to calculate the quantity $\xi_{DL,DW}^j$. To measure the critical current density, we apply a sufficiently large in-plane magnetic field $H_x$ to control the orientation of the in-plane domain-wall spins as described in Ref. [24]. All samples show full switching by a direct current (Figs. 3(a) and 3(c)). The applied $H_x$ was chosen to be slightly larger than the minimum field required for a full switching but not unnecessarily large to significantly reduce the Hall voltage signals. The variation of the required minimum bias field for different samples is due to their different effective Dzyaloshinskii–Moriya-interaction fields. The switching current density was determined from the switching current using a simple parallel-conduction model with measured resistivities for the layers.

Table 1 lists the values of $\xi_{DL,DW}^j$ determined from Eq. [2] and also the ratio $\xi_{DL,DW}^j/\xi_{DL}^j$ to compare to the HHVR values. We find that only for a few samples the two determinations are in semi-quantitative agreement (ratios of 1.1-1.4), but in other cases $\xi_{DL,DW}^j$ can deviate substantially from $\xi_{DL}^j$. $\xi_{DL,DW}^j$ can be unreasonably large, e.g., 1.05 for Au$_{0.25}$Pt$_{0.75}$/Co, 4.0 for W/FeCoB, 2.1 for Ta/Tb$_{20}$Fe$_{64}$Co$_{16}$. The wide variation of the $\xi_{DL,DW}^j/\xi_{DL}^j$ ratio, from 0.12 to 18, indicates that Eq. [2] does not by itself produce a reliable prediction for the SOT switching current in PMA samples.

Here, we note that the discrepancies between $\xi^j_{DL,DW}$ and $\xi^j_{DL}$ cannot be attributed to any effects of Oersted field, fieldlike SOT, in-plane bias field $H_x$, and Joule heating, which are ignored in Eq. [2]. First, the in-plane Oersted field generated by the current flow in the SCG layer is found to have minor effect on the magnetization reversal, e.g. when $j_c$ is as high as $4 \times 10^8$ A/cm$^2$ in Pt 5/Co 1 [42]. The latter indicates an Oersted field, which can be estimated as $j_c d/2$ with $d$ being the layer thickness of the SCG, that is 10-1000 times greater than that for the samples in Table 1. Second, as shown in Table 1, the efficiency of the fieldlike torque ($\xi^j_{FL}$) for the samples we study here shows no obvious correlation to and thus cannot explain the discrepancies between $\xi^j_{DL,DW}$ and $\xi^j_{DL}$. Particularly, when $\xi^j_{FL}$ is zero within the experimental uncertainty, the $\xi^j_{DL,DW}/\xi^j_{DL}$ ratios for Ta 5/Tb$_{20}$Fe$_{64}$Co$_{16}$ 1.8 and Pt 2/Co 1.4 are still as large as 3.5 and 18. Since $\xi^j_{FL}$ of these samples is small, we do not consider this observation to be against the previous report that a *strong* negative fieldlike SOT, which is antiparallel to the Oersted field torque, promoted the switching, whereas a strong positive fieldlike SOT hampered it [42]. Third, it is well known that increasing $H_x$ beyond the minimal required field for a complete switching can lead to increasingly reduced $j_c$ (see the Supplementary Materials [43]) and thus increased $\xi^j_{DL,DW}$ in the domain wall depinning analysis. However, the largest $\xi^j_{DL,DW}/\xi^j_{DL}$ ratios appear for some samples even when $H_x$ is rather small during the switching measurement (e.g. $\xi^j_{DL,DW}/\xi^j_{DL}$ is 10 for W 4/FeCoB 1.5 at 0.1 kOe and 18 for Ta 5/Tb$_{20}$Fe$_{64}$Co$_{16}$ at 0.1 kOe), whereas when $H_x$ is rather large some samples show minimal discrepancies (e.g. 1.1 for Pt 4/Co 0.75 and Pt-Hf/Co 0.63 at $H_x$ = 3 kOe). Under the same $H_x$ of 1.5 kOe, the $\xi^j_{DL,DW}/\xi^j_{DL}$ ratio is 3.2 for Pt 2/Co 1.4 (overestimation) and 0.3 for Pt$_{0.7}$(MgO)$_{0.3}$/Co 0.68 (underestimation). Therefore, we conclude that the effect of in-plane bias field $H_x$ cannot explain these over-/under-estimation of $\xi^j_{DL}$ in the domain wall depinning analysis following Eq. [2]. Fourth, the observed discrepancies between $\xi^j_{DL,DW}$ and $\xi^j_{DL}$ are unlikely to be accounted for by Joule heating, especially for Pt-based HM/Co samples in which $H_k$, Curie temperature, and conductivity are quite high. Note that in the dc switching measurement there is also no substantial reduction of $V_{AH}$ at current densities close to $j_c$ [Fig. 3(b)], while the linear dependence of $H_{DL}$ on $V_{in}$ [see Figs. 2(c) and 2(f)] indicates minimal heating effect in HHVR measurements. In addition, there is not any obvious connection between the discrepancy and the types, coercivity, anisotropy field, thickness, and magnetization of the materials.

Given that in general switching currents should go down when spin-torque efficiencies go up (other factors being equal), it can be tempting to compare switching currents using the simplest possible macrospin approximation (Eq. [1]), even though the reversal is mainly mediated by domain wall depinning. Table 1 lists the estimates of spin torque efficiency $\xi^j_{DL,macro}$ that result from this analysis, as well as the ratios $\xi^j_{DL,macro}/\xi^j_{DL}$. For all of the samples except W 4/Fe$_{0.6}$Co$_{0.2}$B$_{0.2}$ 1.5, $H_k$ is measured from the dependence of the first-harmonic Hall voltage on small in-plane magnetic field $H_x$:

$$V_{1\omega} = \pm V_{AH}\cos\theta \approx \pm V_{AH}(1 - H_x^2/2H_k^2). \qquad (6)$$

Since $V_{1\omega}$ for the W 4/Fe$_{0.6}$Co$_{0.6}$B 1.5 bilayer shows no well-defined parabolic scaling, we determined its $H_k$ value from the in-plane saturation field. The values of $\xi^j_{DL,macro}$ resulting from Eq. [1] are extremely large, varying from 2.0 for Pt 6/Fe$_3$GeTe$_2$ 4, to 234 for Ta 5/TbFeCo 1.8, and to 306 for W 4/Fe$_{0.6}$Co$_{0.2}$B$_{0.2}$ 1.5, thereby overestimating the true $\xi^j_{DL}$ values by a factor of 10-10$^3$. This failure is no doubt due to the inapplicability of the macrospin approximation in the micrometer-sized SCG/FM samples, and is consistent with previous reports that $j_c$ increases as the FM size decreases [44] and is much smaller than expected for macrospin reversal when the FM size is > 80 nm [44-46]. $\xi^j_{DL,macro}$ does not even provide reliable guidance about the relative values of $\xi^j_{DL}$ for different materials, as the ratio $\xi^j_{DL,macro}/\xi^j_{DL}$ varies between samples by more than a factor of 180 (and by more than a factor of 5 even among the just samples with FM layers of Co).

Finally, we note that even in the absence of any interpretive model there is no clear inverse correlation between $j_c$ and $\xi^j_{DL}$ when comparing heterostructures made from different materials. For example, $j_c$ for the W 4/FeCoB 1.5 sample ($\xi^j_{DL} = 0.4$) is $3.6 \times 10^5$ A/cm$^2$, which is close to that for Ta 5/Tb$_{20}$Fe$_{64}$Co$_{16}$ 1.8 ($\xi^j_{DL} = 0.12$, Table 1), despite the more than threefold difference in $\xi^j_{DL}$. Even limiting the comparison to the SCG/Co samples in Table 1, the product of $j_c\xi^j_{DL}$ varies by a factor of almost 5. We conclude that comparisons of current densities for switching, whether by themselves or when analyzed within the framework either a domain-wall model of Eq. [2] or a macrospin model (Eq. [1]), do not provide reliable qualitative information about the relative values of torque efficiencies or the priority of the SCGs.

In drawing this conclusion we do not wish to imply that the domain-wall effective-field picture of Eq. [2] is not useful in other circumstances. We expect that in the careful procedure described in ref. [26], in which small current-induced shifts in the out-of-plane switching field are measured in the presence of an in-plane field sufficient to orient spins within the domain walls, the domain wall picture is appropriate to describe the relatively small shifts. This is supported by the fact that the hysteresis loop shift measurements [26] reported $\xi^j_{DL} = 0.15$ for Pt/Co and $\xi^j_{DL} = 0.12$ for Ta/FeCoB, both of which are consistent with the HHVR measurements (Table 1). However, it appears that for predicting the full switching current of micron-scale PMA samples, the SOT is not simply equivalent to an out-of-plane magnetic field that depends only on the saturation magnetization, as Eq. [2]. The physics at work might also depend on other material factors, including the Dzyaloshinskii–Moriya interaction and/or the magnetic damping [47], which requires future efforts to understand.

In conclusion, we have demonstrated by a variety of examples that there is no simple correlation between the critical switching current density $j_c$ and $\xi_{DL}^j$ of a micro-sized perpendicular SCG/FM bilayer. As a consequence, the magnitudes of $j_c$ by themselves do not provide reliable guidance about the relative strengths of $\xi_{DL}^j$ or the relative potential of different spin-current-generation materials for technological applications. We find that $\xi_{DL}^j$ can be overestimated by up to thousands of times if $j_c$ is assumed to represent the critical switching current density of a rigid macrospin. When $j_c$ is assumed as the critical current for depinning chiral domain walls, $\xi_{DL}^j$ can be either under-estimated or overestimated by up to tens of times. As more reliable means of evaluating $\xi_{DL}^j$, we recommend out-of-plane HHVR measurements of strong PMA samples [33-37], current-dependent shifts in out-of-plane switching field in PMA samples according to the careful procedure of ref. [26], angle-dependent HHVR measurements of in-plane anisotropy samples [33-37], or antidamping switching measurements of in-plane magnetized 100-nm scale magnets [4,35]. The FM thickness-dependent spin-torque ferromagnetic resonance [48,49] can also be used, but appears to yield slightly smaller values of $\xi_{DL}^j$ values than HHVR measurements [50,51]. Our results should benefit the understanding of SOT evaluation and the perpendicular magnetization switching process.

This work was supported in part by the Office of Naval Research (N00014-15-1-2449), in part by the Defense Advanced Research Projects Agency (USDI D18AC00009), in part by the NSF MRSEC program (DMR-1719875) through the Cornell Center for Materials Research, and in part by the NSF (ECCS-1542081) through use of the Cornell Nanofabrication Facility/National Nanotechnology Coordinated Infrastructure.

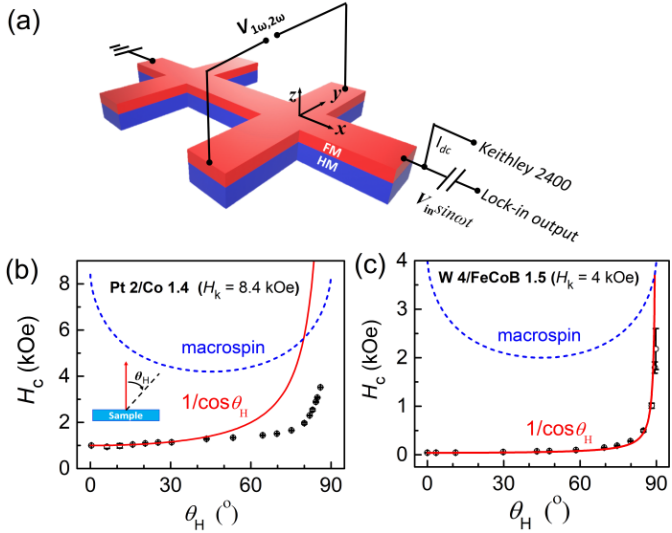

Fig. 1. (a) Geometry and coordinates for HHVR and switching measurements. $H_c$ vs $\theta_H$ for (b) Pt 2/Co 1.4 and (c) W 4/FeCoB 1.5, showing a significant deviation from the expectation of macrospin [$H_c = H_k (\cos^{2/3}\theta_H + \sin^{2/3}\theta_H)^{-3/2}$, dashed blue line], while is relatively consistent with domain wall depinning ($\propto 1/\cos\theta_H$, solid red line). The $H_c$ values are determined from $V_{1\omega}$ hysteresis loops ($V_{in} = 0.1\ V$) with the external field swept in the $xz$ plane at different fixed polar angle of $\theta_H$.

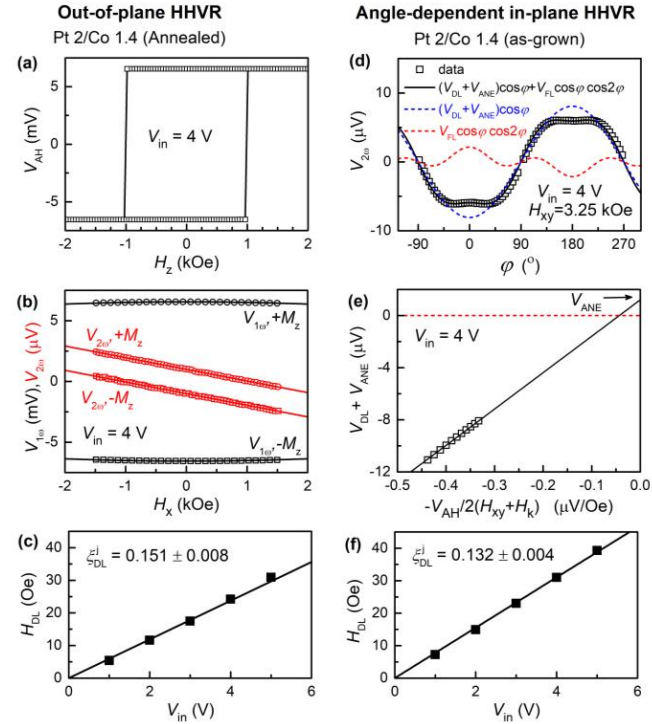

Fig. 2. Harmonic Hall voltage response measurements. (a) $V_{AH}$ vs $H_z$, (b) $V_{1\omega}$ and $V_{2\omega}$ vs $H_x$, (c) $H_{DL}$ vs $V_{in}$ for perpendicularly magnetized Pt 2/Co 1.4 (annealed). (d) $V_{2\omega}$ vs $\varphi$, (e) $V_{DL}$ vs -$V_{AH}/2(H_{xy}+H_k)$, (f) $H_{DL}$ vs $V_{in}$ for in-plane magnetized Pt 2/Co 1.4 (as-grown). The solid straight lines in (b), (c), (e), and (f) represent the best linear fits; the solid parabolic lines in (b) represent the best fits of data to Eq. [6].

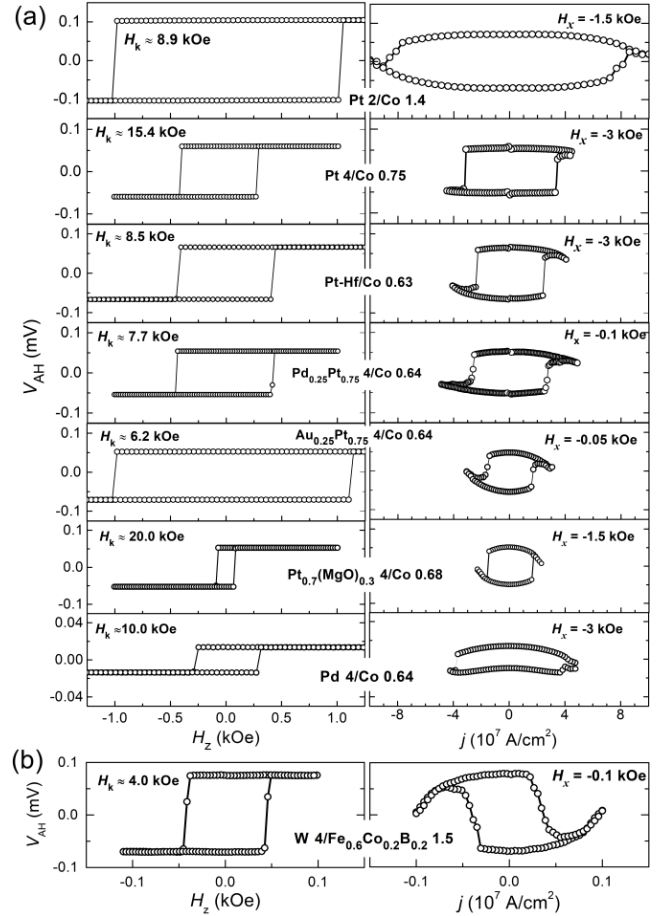

Fig. 3. (a) $V_{AH}$ vs out-of-plane field ($H_z$) and $V_{AH}$ vs dc current density inside the HM layer ($j$) for perpendicularly magnetized Pt 2/Co 1.4, Pt 4/Co 0.63, Pt-Hf/Co 0.75, Pt$_{0.75}$Pd$_{0.25}$ 4/Co 0.64, Au$_{0.25}$Pt$_{0.75}$ 4/Co 0.64, Pt$_{0.7}$(MgO)$_{0.3}$ 4/Co 0.68, and Pd 4/Co 0.64, respectively. (b) $V_{AH}$ vs $H_z$ and $V_{AH}$ vs $j$ for perpendicularly magnetized W 4/Fe$_{0.6}$Co$_{0.2}$B$_{0.2}$ 1.5. For both field and current switching, the applied sinusoidal voltage has a magnitude of $V_{in} = 0.1\ V$. Here, the Pt 2/Co 1.4 shows a reduced $V_{AH}/V_{in}$ ratio compared to that of Fig. 2(a) due to the additional "annealing" in the current switching measurement where the applied direct current density was up to $1.2 \times 10^8$ A/cm$^2$, which is very high. However, we find this additional annealing has no significant influence on the coercivity and the SOT efficiency.

Table 1. Comparison of spin-torque efficiencies determined from harmonic response ($\xi_{DL}^j$) and magnetization switching ($\xi_{DL,DW}^j$, $\xi_{DL,macro}^j$) of perpendicularly magnetized bilayers. The values of $\xi_{DL,DW}^j$ are determined from a model of a current-induce effective field acting on domain walls (Eq. [2]), and $\xi_{DL,macro}^j$ is determined within a macrospin model (Eq. [1]). Layer thicknesses are in nm. $\rho_{xx}$ is the resistivity of the heavy metal layer, $M_s$, $H_c$, $H_k$, and $j_{c0}$ are the magnetization, the coercivity, the perpendicular anisotropy field, the switching current density of the perpendicular magnetic layer. $\xi_{FL}^j$ is the fieldlike spin-orbit torque efficiency determined from harmonic response measurements. The $\xi_{DL}^j$ and $\xi_{FL}^j$ results for the Pt 6/Fe$_3$GeTe$_2$ 4 and Ta 5/Tb$_{20}$Fe$_{64}$Co$_{16}$ 1.8 samples were reported in ref. [40] and [41], while we calculate the corresponding values of $\xi_{DL,DW}^j$ and $\xi_{DL,macro}^j$ are calculated in this work using the reported $j_c$ values and other sample parameters as reported in [40] and [41].

| Samples | $\rho_{xx}$ ($\mu\Omega$ cm) | $M_s$ (emu/cm$^3$) | $H_k$ (kOe) | $H_C$ (kOe) | $H_x$ (kOe) | $j_c$ (10$^7$ A/cm$^2$) | $\xi_{DL}^j$ | $\xi_{DL,DW}^j$ | $\xi_{DL,macro}^j$ | $\xi_{DL,DW}^j/\xi_{DL}^j$ | $\xi_{DL,macro}^j/\xi_{DL}^j$ | $\xi_{FL}^j$ |
|---|---|---|---|---|---|---|---|---|---|---|---|---|
| Pt 2/Co 1.4 (annealed) | 67 | 1454 | 8.9 | 1.0 | 1.5 | 8.2 | 0.15 | 0.48 | 2.8 | 3.2 | 18.7 | -0.003 |
| Pt 4/Co 0.75 | 51 | 1450 | 14.8 | 0.35 | 3.0 | 3.2 | 0.21 | 0.23 | 6.0 | 1.1 | 28.6 | -0.049 |
| Pt-Hf/Co 0.63 | 140 | 1720 | 8.5 | 0.43 | 3.0 | 2.4 | 0.36 | 0.38 | 3.8 | 1.1 | 10.6 | -0.013 |
| Pt$_{0.75}$Pd$_{0.25}$ 4/Co 0.64 | 58 | 1706 | 7.7 | 0.44 | 0.1 | 2.6 | 0.26 | 0.36 | 4.8 | 1.4 | 18.5 | -0.059 |
| Au$_{0.25}$Pt$_{0.75}$ 4/Co 0.64 | 84 | 1412 | 6.4 | 1.0 | 0.05 | 1.7 | 0.30 | 1.05 | 5.2 | 3.5 | 17.3 | -0.116 |
| Pt$_{0.7}$(MgO)$_{0.3}$ 4/Co 0.68 | 71 | 1300 | 20.0 | 0.08 | 1.5 | 1.5 | 0.30 | 0.09 | 16.8 | 0.3 | 56 | -0.039 |
| Pd 4/Co 0.64 | 40 | 1660 | 10.0 | 0.17 | 3.0 | 3.75 | 0.07 | 0.09 | 3.0 | 1.3 | 42.9 | -0.054 |
| W 4/Fe$_{0.6}$Co$_{0.2}$B$_{0.2}$ 1.5 | 185 | 1240 | 4.0 | 0.04 | 0.1 | 0.036 | 0.4 | 4.0 | 306 | 10 | 765 | N. A. |
| Pt 6/Fe$_3$GeTe$_2$ 4 | 75 | 16 | 25 | 0.62 | 0.5 | 1.2 | 0.12 | 0.06 | 2.0 | 0.5 | 16.7 | 0.054 |
| Ta 5/Tb$_{20}$Fe$_{64}$Co$_{16}$ 1.8 | 288 | 350 | 10 | 0.07 | 0.2 | 0.04 | 0.12 | 2.1 | 234 | 18 | 1950 | 0 |

# Lack of simple correlation between switching current density and spin-orbit torque efficiency of perpendicularly magnetized spin-current generator/ferromagnet heterostructures

Lijun Zhu[1*], D. C. Ralph[1,2], R. A. Buhrman[1]
1. Cornell University, Ithaca, New York 14850, USA
2. Kavli Institute at Cornell, Ithaca, New York 14850, USA

*lz442@cornell.edu

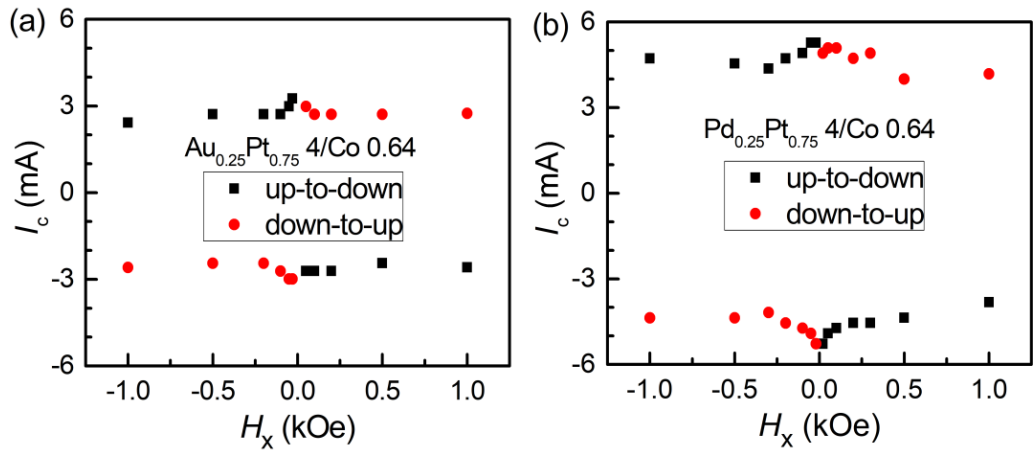

Fig. S1. In-plane magnetic field dependence of switching current for (a) $Au_{0.25}Pt_{0.75}$ 4 nm/Co 0.64 nm and (b) $Pd_{0.25}Pt_{0.75}$ 4 nm/Co 0.64 nm.